# A physical perspective to understand myelin. I. A physical answer to Peter's quadrant mystery

**Authors**: Yonghong Liu[1], Yapeng Zhang[1], Wenji Yue[1], Ran Zhu[1], Tianruo Guo[3], Fenglin Liu[1], Yubin Huang[1], Tianzhun Wu*[1,2], Hao Wang*[1,2]

**Affiliations:**

[1]Institute of Biomedical & Health Engineering, Shenzhen Institutes of Advanced Technology (SIAT), Chinese Academy of Sciences (CAS), Shenzhen 518035, China

[2]Key Laboratory of Health Bioinformatics, Chinese Academy of Sciences

[3]Graduate School of Biomedical Engineering, University of New South Wales, Sydney, NSW2052, Australia

***Hao Wang** hao.wang@siat.ac.cn
***Tianzhun Wu** tz.wu@siat.ac.cn

**Abstract**

In the development of oligodendrocytes in the central nervous systems, the inner and outer tongue of the myelin sheath tend to be located within the same quadrant, which was named as Peters quadrant mystery. In this study, we conduct *in silico* investigations to explore the possible mechanisms underlying the Peters quadrant mystery. A biophysically detailed model of oligodendrocytes was used to simulate the effect of the actional potential-induced electric field across the myelin sheath. Our simulation suggests that the paranodal channel connecting the inner and outer tongue forms a low impedance route, inducing two high-current zones at the area around the inner and outer tongue. When the inner tongue and outer tongue are located within the same quadrant, the interaction of these two high-current-zones will induce a maximum amplitude and a polarity reverse of the voltage upon the inner tongue, resulting in the same quadrant phenomenon. This model indicates that the growth of myelin follows a simple principle: an external negative or positive E-field can promote or inhibit the growth of the inner tongue, respectively.

**Introduction**
Since the pioneering electron microscope (SE) observations of the spiral structure of myelin sheath were conducted between the 1950s and 1980s [1-2], the ultrastructure and function of the myelin sheath have been paid more attention in neuroscience [3][4]. The myelin sheath was initially reported as a pure electrical insulator, enabling a "saltatory" impulse propagation [5]. However, this hypothesis cannot explain many experimental observations in myelin ultrastructures. For example, myelin in the superficial layers of the cortex has diversified longitudinal distribution [6]; myelin sheaths in the peripheral nervous system (PNS) spiral oppositely and the same to its neighbor on the same [7][8] and adjacent axon, respectively [9]; and in particular, Peters quadrant mystery [10,11,14-19]-has been observed in many myelin sheaths. These non-trivial ultrastructures imply that the function of the myelin is more than an insulating layer.

An anatomically accurate and biophysically detailed model can improve our understanding of myelin ultrastructures and functions. For example, a coil inductor model of the spiraling structure was used to understand the unique spiraling directions between adjacent myelin sheaths. [12] To achieve a positive mutual inductance, the neighboring myelin on the same axon shall have opposite spiraling directions, while the neighboring myelin on the adjacent axons shall have the same spiraling direction. This simulation has been confirmed by SEM observations [11].

This study follows the same research paradigm to explore the possible mechanisms underlying the Peters quadrant mystery. In particular, the myelin sheath is modeled as a distributed parameter circuit [13], and the electric field (E-field) distribution induced by neural electric activities is investigated *in silico*. The simulated E-field was used to explain why the inner tongue and outer tongue of the myelin sheath tend to locate in the same quadrant, a repeatedly observed intriguing phenomenon [10,11,14-19]. The new knowledge gained in this study provides new insights into the relationships between neural electrical activity and myelin growth.

**Peters quadrant mystery**
During axon growth, the myelin wraps around it as a spiral "bandage." However, there is an interesting tendency for this spiral's initial and endpoints to occur close together, as if the myelin were insisting on running only complete laps of the arena [14] [15]. This is analogous to winding rope into a film spool until the rope spills at the angle where the initial "lump" occurs. Initial and endpoints will tend to occur within the same "quadrant" (Figure1(a)). Peters first observed this phenomenon in the optic nerves of rodent models in 1964, then further confirmed by multiple studies in visual callosal [11], dorsal and anterior root axon [16][17], and sural nerves [18][19]. Interestingly, Schwann cell myelination in PNS demonstrated quadrant tendency diminishing gradually with the thickening of myelin [17]. In contrast, myelination of oligodendrocytes in the CNS exhibits a stronger tendency with the thickening of myelin [10].

The actual experimental result of Peter's quadrant mystery is illustrated in Figure 1(a&b) with the reproduced data from Peter's observation (Figure 1(b)) [10]. The angle between the outer tongue and the inner tongue is defined as β. As seen in Figure 1(b), the occurrence probability of the case when β is within the first quadrant (each quadrant is 45º) is 52.2%, which is much higher than other quadrants. There is an abrupt change of the probability when the inner tongue grows from the last quadrant to the first quadrant.

Peter's quadrant mystery indicates two points:
1. The outer tongue exerts an effect upon the growth rate of the inner tongue;
2. This effect has an abrupt change when the inner tongue goes through the radial quadrant where the outer tongue is located.

How can the outer tongue affect the growth rate of the inner tongue? This is still an open question that does not seem to have a biological answer (like protein, molecule, and gene) [15]. Thus, a physical hypothesis is proposed to build the connection between the growth rate of the inner tongue and the position of the outer tongue. We assume that the electric field (E-field) upon the inner tongue can modulate the inner tongue's growth rate. A distributed parameter circuit modeling the cross-section of the myelin sheath is built to analyze this E-field.

**Method**

1. **The circuit simulation**

In this study, a cross-section of a myelinated axon is modeled as a distributed parameter circuit, as shown in Figure 1(d). The transmembrane parts can be modeled as an RC circuit, while the non-transmembrane parts are modeled as resistors. It is emphasized that the inner tongue and outer tongue are connected with a paranodal channel (Figure 1(c)), which is a path filled with cytoplasmic liquid and forms a low impedance route. Therefore, in terms of the circuit, the outer tongue and inner tongue are connected with a resistor, whose impedance is low. This low impedance route is critical for the simulation. Since the inner tongue is the growing terminal, the transmembrane voltage of the inner tongue is measured in the simulation, as shown in Figure 1(e). The inside terminal of the inner tongue is set as the reference during the measurement.

2. **The origin of the current source**

The origin of the current source implemented in the circuit simulation is the action potential. The explanation of the waveform and polarity of the current source is illustrated in Figure 2. Figure 2(a) shows a typical waveform of the action potential. Since we only consider the absolute voltage change (start from 0 mV rather than -70 mV), the action potential is very similar to a positive monophasic voltage waveform (take the inside terminal of the axon as the reference point in Figure 1(d)). When an inward current happens at the Ranvier node (the depolarization phase with inward $Na^+$ ionic current of the action potential), an outward current will happen across the internode (Figure 2(b)). Therefore, the E-field across the myelin has a dominant positive component, which is equivalent to a current from the inside to the outside. So the current source applied in Figure 1(d) has a positive monophasic current waveform with its positive terminal connected with the inside terminal of the axon. A more detailed modeling approach and circuit parameter selection can be found in **Supplementary S1-S2**.

## Result

### 1. The current is more concentrated in the inner-tongue-zone and outer-tongue-zone

The amplitude of the current across each transmembrane capacitor in Figure 1(d) is recorded and re-distributed into the round shape analogous to the circle of the myelin sheath, as shown in Figure 3(a). The current emitted from the axon is not distributed radially identical. Instead, the current is preferentially concentrated at the area close to the inner tongue and outer tongue, called inner-tongue-zone and outer-tongue-zone, respectively (Figure 3(a)). As shown in Figure 3(b), these high current zones move with the position of the inner tongue and outer tongue, showing that these two high current zones are directly induced by the existence of the inner and outer tongues. The cause of these two high current zones is qualitatively explained in Figure 3(c).

Since the current source is connected with the inner terminal and outer terminal, it can be considered that the current is emitted from the axon to the outer space of the myelin sheath. The current emission follows a specific distribution, as shown in Figure 3(c-i). This distribution can be understood by a two-step analysis shown in Figure 3(c-ii&c-iii). Firstly, due to the existence of the low impedance route, the inner and outer tongue can concentrate more current, shown in Figure 3(c-ii):

$$I_{inner} \gg I_{radial} \ \& \ I_{outer} \gg I_{radial};$$

The current into the inner tongue is shared between the radial path and the channel path, shown in Figure 3(c-iii):

$$I_{inner} = I_{inner-radial} + I_{inner-channel};$$

It is the same for the current into the outer tongue:

$$I_{outer} = I_{outer-radial} + I_{outer-channel};$$

Although $I_{inner-radial}$ is just part of the total current into the inner tongue, due to the current concentration effect, it still can be concluded that:

$$I_{inner-radial} > I_{radial};$$

In general, due to the current concentration effect by the low impedance route connecting the inner and outer tongues, the radial current at the area close to the inner and outer tongue will be higher than that of the other position, resulting in the two high current zones. This is an intuitive and qualitative explanation. A more detailed quantitative explanation based on circuit analysis can be found in **Supplementary S2**. All model details and parameter values are described in **Table 1** in **Supplementary S1**.

### 2. The radial angle influences the electric voltage on the inner tongue

Since the inner tongue is the growing terminal, we will focus on how the voltage changes on the inner tongue, which is the voltage measured on the specific capacitor representing the tip membrane of the inner tongue shown in Figure 1(e). As explained in Figure 2, the action potential can be approximately considered as a monophasic positive current pulse. The actual applied current waveform in the simulation is also a monophasic current pulse shown in Figure 4(a). During the growth of the inner tongue, the voltage amplitude on the inner tongue will have a periodical maximum when the inner-tongue-zone is radially overlapped with the outer-tongue zone (located at the same quadrant in Figure 4(b)). When the inner tongue is located at position 1 (Figure 4(b-i)), the voltage waveform as a maximum positive peak. When it is located at position 2 (Figure 4(b-ii)), the voltage waveform has a maximum negative peak. As seen, the voltage upon the inner tongue has a polarity reverse when it grows from position 1 to position 2. A continuous change of the maximum voltage with the inner tongue growth is shown in Figure 4(c), showing a periodic polarity reverse. As seen, the transmembrane voltage of the inner tongue is affected by the relative position between the inner tongue and outer tongue (or the radial angle between the inner and outer tongue). Meanwhile, it has an abrupt change when the inner and outer tongues are in the same quadrant. If the growth rate of the inner tongue is modulated by the polarity and amplitude of this voltage, the same quadrant mystery can have a simple answer, as explained below.

### 3. The voltage polarity reverse

An illustrative drawing to explain the polarity reverse is shown in Figure 5. When the inner tongue and the outer tongue are located at the same quadrant, the two high-current zones will have interaction, forming a directional current flow from the inner tongue towards the outer tongue (Figure 5(a)). When the inner tongue is located at position 1, the radial current of the inner tongue, $I_{inner-radial}$, towards the position of the outer tongue will form a transmembrane current upon the inner tongue with an outward direction shown in Figure 5(b-i), which is equivalent to an externally applied negative E-field. When the inner tongue is located at position 2, this transmembrane current has an inward direction, which is opposite to the situation of position 1 (Figure 5(b-ii)) and is equivalent to an externally applied positive E-field. This is the reason for the polarity reversed in Figure 4.

### 4. The explanation to Peter quadrant mystery

The relationship between the transmembrane voltage of the inner tongue and its radial position is illustrated in Figure 6. It is emphasized that the curve in Figure 6 is an illustrative drawing, not an accurate duplication from the simulation results. When the inner tongue locates in position 1 (entering the outer-tongue-zone), the transmembrane E-field of the inner tongue reaches the maximum outward value. In Peters's observations, position 1 showed the lowest occurrence frequency [10], indicating the fastest growth rate. With further growth, the inner tongue will reach position 2 (leaving the outer-tongue-zone). The transmembrane E-field of the inner tongue reaches the maximum inward value. The occurrence frequency at this position is the highest, indicating the slowest growth rate. Therefore, we can conclude that the growth rate is correlated with the polarity (direction) and amplitude of the transmembrane E-field. An outward E-field can facilitate growth, while an inward E-field can inhibit growth. In other words, an extracellular negative E-field can promote myelin growth, while an extracellular positive E-field can inhibit myelin growth.

Our simulations suggested that the existence of the paranodal channel connecting the inner and outer tongue forms two high-current zones. When these two high-current-zones are getting close, which happens when the inner and outer tongues are located within the same quadrant, the voltage upon the inner tongue will have a maximum amplitude and a polarity reverse, resulting in a minimum growth rate at position 2. So the inner tongue tends to stay at position 2, observed as the same quadrant phenomenon. Interestingly, this phenomenon does not only appear in Oligodendrocytes [10][11] but also exists in the early stage of the myelination by Schwann cells in PNS [17]. It is known that for a mature Schwann cell, its inner tongue, and outer tongue form two radial circles shown in Figure 7, rather than just occupying a certain radial angle, which is the situation of Oligodendrocytes. Therefore, the same quadrant tendency will diminish for a mature Schwann cell [20].

**5. A possible explanation for g-ratio**

The g-ratio is the ratio of the inner axonal diameter to the total outer diameter, including the myelin sheath [21]. The g-ratio ranged from 0.72 to 0.81 in CNS, and 0.46 to 0.8 in PNS. However, if the axon diameter is less than 0.4 μm, it will fail to form the myelin sheath [22], indicating the key role of the axonal physical properties in terminating the growth of myelin. Although there are still controversies [22][23], earlier studies suggested the contribution of g-ratio in modulating conduction velocity [24]. However, this theory fails to build the connection between the signal propagation and the inner tongue, which is the growth terminal of the myelin. In this study, the modeled E-field tends to decrease by increasing the layers of myelin layers. Thus, when the E-field is lower than a certain threshold, the growth of the inner tongue will be automatically terminated. Thus, this theory indicates the potential correlation between the g-ratio and the E-field on the inner tongue. In our next study, a more detailed model is proposed to explain the g-ratio phenomenon [28].

**Conclusion**

The physical origin of the same quadrant mystery is the preferential E-field distribution on the cross-section of the myelin. Since actional potentials induce E-field, it explains the relation between neural electric activity and the ultrastructure of myelin. Furthermore, the preferential E-field distribution resulting from the breaking of the central symmetry by the outer tongue explains the difference of the "same quadrant" observation between Oligodendrocytes in CNS and Schwann cells in PNS. Meanwhile, this study also reveals the physical factor that modulates myelin growth: an extracellular negative or positive E-field can promote or inhibit myelin growth, respectively. Finally, the computational approach can probe neuronal ultrastructures at a resolution far beyond the current state-of-the-art biological experiments, providing a promising tool to explore neuroscience from a physical perspective.

**Acknowledgments**

This work was supported by the grant from Guangdong Research Program (2019A1515110843), Shenzhen Research Program (JCYJ20170818152810899, GJHZ20200731095206018), Chinese Academy of Sciences Research Program (2011DP173015,172644KYSB20190077) and National Natural Science Foundation of China grants(31900684).

**Reference**
1. BEN GEREN B. The formation from the Schwann cell surface of myelin in the peripheral nerves of chick embryos. *Exp Cell Res*, 1954, 7(2):558-62. doi: 10.1016/s0014-4827(54)80098-x. PMID: 13220597.
2. Bunge RP, Bunge MB, Bates M. Movements of the Schwann cell nucleus implicate progression of the inner (axon-related) Schwann cell process during myelination. *J Cell Biol*, 109: 273–284, 1989. doi:10.1083/jcb.109.1.273.
3. Chang K J, Redmond S A, Chan J R. Remodeling myelination: implications for mechanisms of neural plasticity. *Nat Neurosci*, 2016, 19(2): 190-197.
4. Monje M. Myelin plasticity and nervous system function. *Annu Rev Neurosci*,, 2018, 41: 61-76.
5. Boullerne AI. The history of myelin. *Exp Neurol*, 283, Pt B: 431–445, 2016. doi:10.1016/j.expneurol.2016.06.005.
6. Tomassy GS, Berger DR, Chen HH, Kasthuri N, Hayworth KJ, Vercelli A, Seung HS, Lichtman JW, Arlotta P. Distinct profiles of myelin distribution along single axons of pyramidal neurons in the neocortex. *Science*, 2014, 344(6181):319-24. doi: 10.1126/science.1249766. PMID: 24744380; PMCID: PMC4122120.
7. Uzman, BG, and Nogueira-Graf, G. Electron microscope studies of the formation of nodes of ranvier in mouse sciatic nerves. *J Cell Biol*, 1957, 3, 589–598. doi: 10.1083/jcb.3.4.589
8. Armati, P J, and Mathey, E K. An update on schwann cell biology—immunomodulation, neural regulation and other surprises. *J Neurol Sci*, 2013, 333, 68–72. doi: 10.1016/j.jns.2013.01.018
9. Richards W, Kalil R, Moore CL. An observation about myelination. *Exp Brain Res*, 1983, 52(2):219-25. doi: 10.1007/BF00236630. PMID: 6641884.
10. Peters A. Further observations on the structure of myelin sheaths in the central nervous system. *J Cell Biol*, 1964, 20(2): 281-296.
11. Waxman S G, Swadlow H A. Ultrastructure of visual callosal axons in the rabbit. *Exp Neurol.* , 1976, 53(1): 115-127.
12. Wang H, Wang J, Cai G, et al. A Physical Perspective to the Inductive Function of Myelin—A Missing Piece of Neuroscience. *Front Neural Circuits*, 2021, 14: 86.
13. Luo, J, Zhang, K, Chen, T, Zhao, G, Wang, P and Feng, S. ,Distributed parameter circuit model for transmission line. In 2011 International Conference on Advanced Power System Automation and Protection, 2011, 2:1529-1534.
14. Webster HD. The geometry of peripheral myelin sheaths during their formation and growth in rat sciatic nerves. *J Cell Biol*, 1971, 48(2):348-67. doi: 10.1083/jcb.48.2.348. PMID: 4928020; PMCID: PMC2108190.
15. Traill RR. Strange regularities in the geometry of myelin nerve-insulation—a possible single cause. *Ondwelle short-monograph*, 2005, 1(1): 1-9.
16. Berthold CH, Carlstedt T. Myelination of S1 dorsal root axons in the cat. *J Comp Neurol*,1982, 209(3): 225-232.
17. Fraher JP. A quantitative study of anterior root fibres during early myelination. *J Anat.*, 1972, 112(1): 99.

18. Schröder JM. Developmental and pathological changes at the node and paranode in human sural nerves. *Microsc Res Tech*, 1996 , 34(5):422-35. doi: 10.1002/(SICI)1097-0029(19960801)34:5<422::AID-JEMT2>3.0.CO;2-O. PMID: 8837018.
19. Bertram M, Schröder J M. Developmental changes at the node and paranode in human sural nerves: morphometric and fine-structural evaluation. *Cell Tissue Res.*, 1993, 273(3): 499-509.
20. Gomez-Sanchez J A, Pilch K S, van der Lans M, et al. After nerve injury, lineage tracing shows that myelin and Remak Schwann cells elongate extensively and branch to form repair Schwann cells, which shorten radically on remyelination. *J Neurosci*, 2017, 37(37): 9086-9099.
21. Chomiak T, Hu B. What is the optimal value of the g-ratio for myelinated fibers in the rat CNS? A theoretical approach. *PLoS One*, 2009, 4(11):e7754. doi: 10.1371/journal.pone.0007754. PMID: 19915661; PMCID: PMC2771903.
22. Waxman SG, Bennett MV. Relative conduction velocities of small myelinated and non-myelinated fibres in the central nervous system. *Nat New Biol*, 1972, 238(85):217-9. doi: 10.1038/newbio238217a0. PMID: 4506206.
23. Smith R S, Koles Z J. Myelinated nerve fibers: computed effect of myelin thickness on conduction velocity. Am. J. Physiol, 1970, 219(5): 1256-1258.
24. RUSHTON WA. A theory of the effects of fibre size in medullated nerve. *J Physiol*, 1951, 115(1):101-122. doi:10.1113/jphysiol.1951.sp004655
25. Sukhorukov VL, et al. Hypotonically induced changes in the plasma membrane of cultured mammalian cells. *J Membr Biol.* 1993, 132(1):27-40. doi: 10.1007/BF00233049. PMID: 8459447.
26. Goldup, A., Ohki, S. and Danielli, J.F., 1970. Black lipid films. *In Recent progress in surface science* (Vol. 3, pp. 193-260). Elsevier.
27. Cole, K.S., 1941. Rectification and inductance in the squid giant axon. *Journal of General Physiology*, 25(1), pp.29-51.
28. Liu, Y., Zhang, Y., Yue, W., Zhu, R., Guo, T., Liu, F., Wu, T. and Wang, H., 2021. A Physical perspective to understand the mechanism of myelin development. arXiv preprint arXiv:2111.13689.

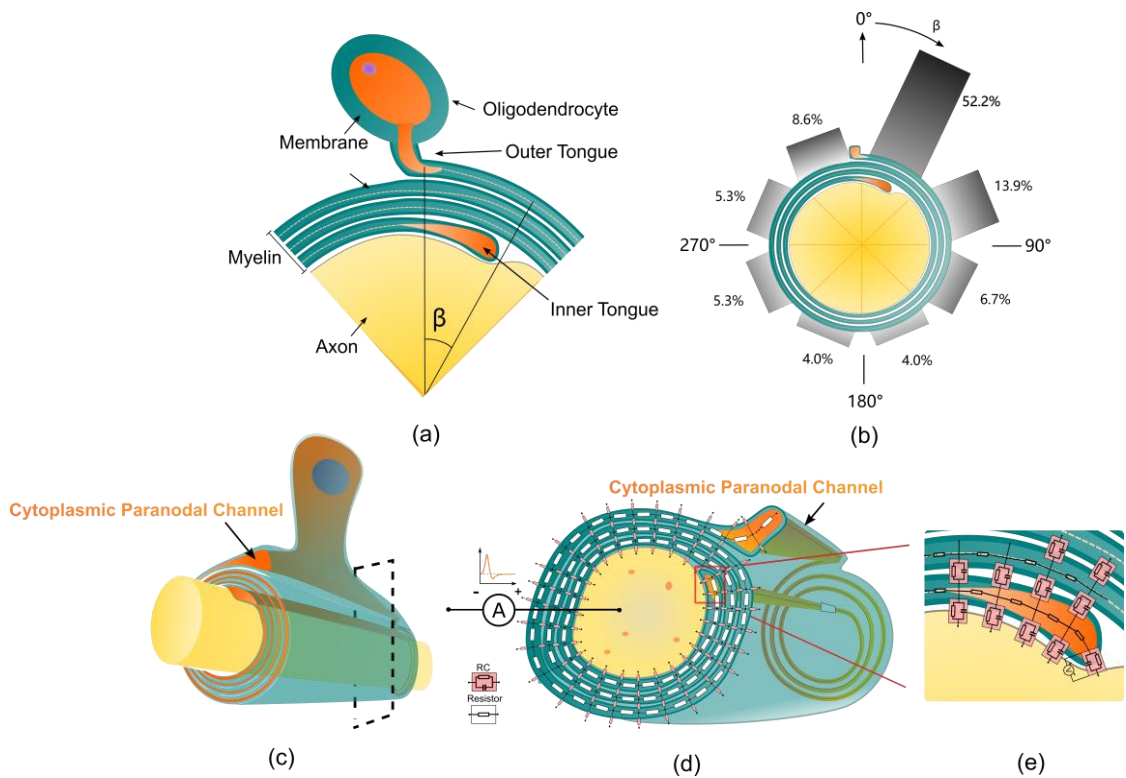

Figure 1. (a) The illustration of the Peters quadrant mystery in Oligodendrocytes: there is a relative angle β between the inner and outer tongue. When β is lower than 50º, the inner and outer tongues are considered in the same quadrant. (b) The frequency of β within each 45º octant (Reproduced from Peters observation [10]). (c) An illustrative drawing of an axon myelinated by an oligodendrocyte, the orange part indicates the cytoplasmic paranodal channel connecting the inner and outer tongue; (d) The equivalently distributed circuit network model of the cross-section of a myelinated axon. Two kinds of circuit components representing different local electrical properties of the myelin sheath are shown. (d) The transmembrane capacitance of the growth terminal of the inner tongue.

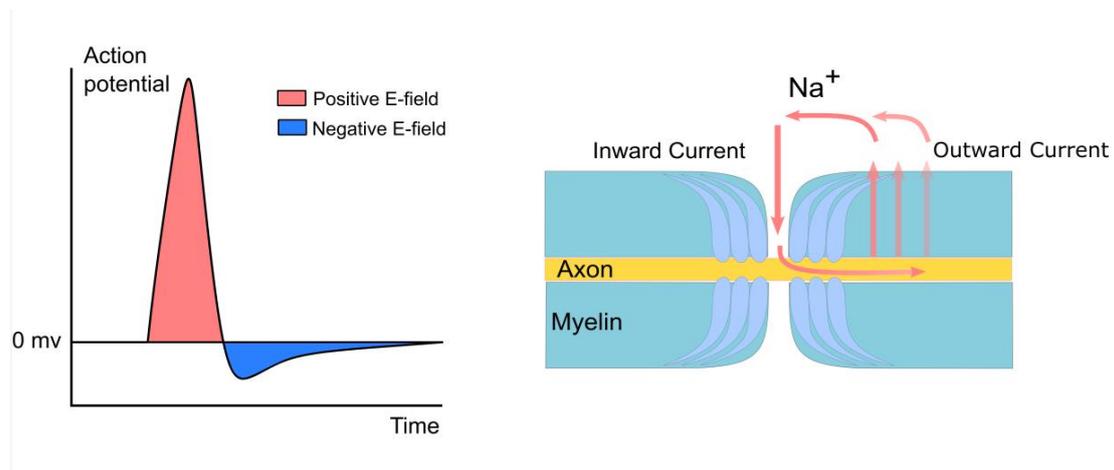

Figure 2. A typical waveform of the action potential and current path between the node of Ranvier and internode of myelin.

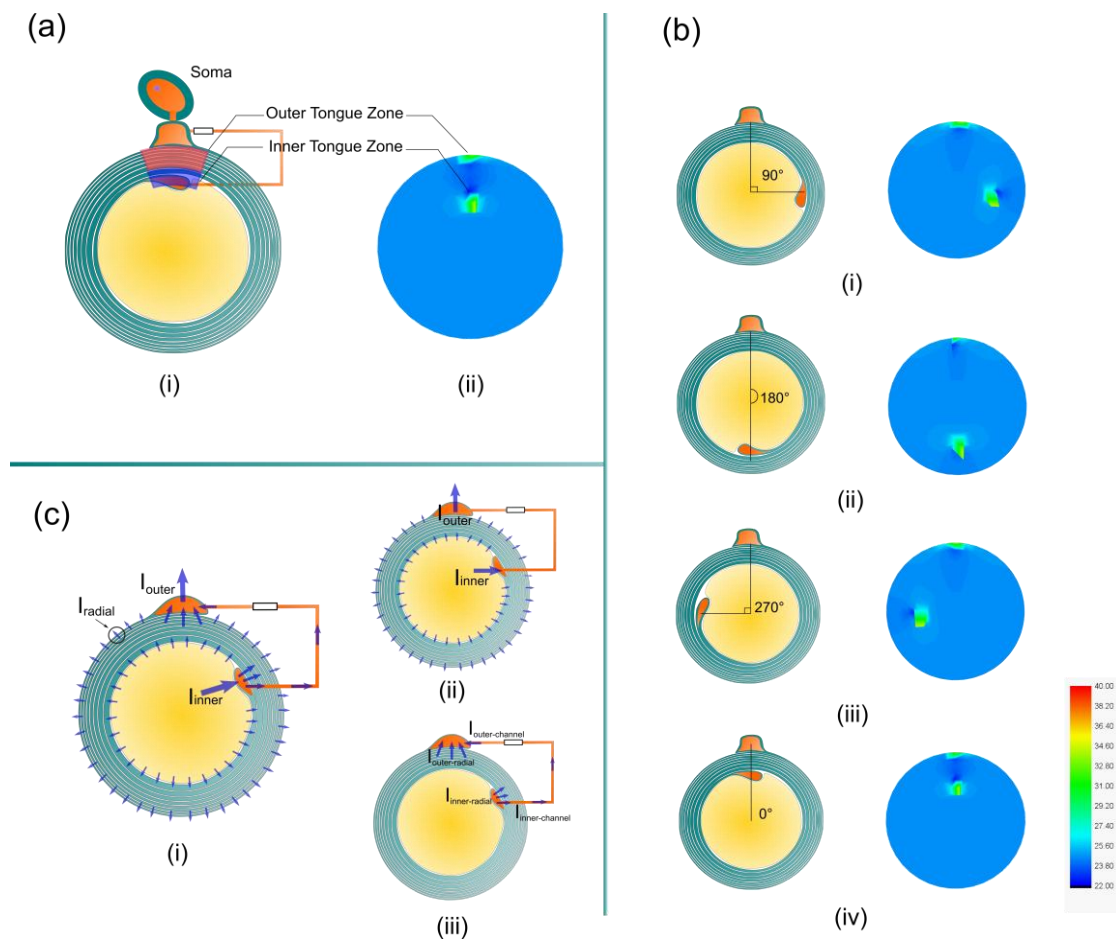

Figure 3. (a) The high current zones related to the position of the inner tongue and the outer tongue in the simulation is named as inner-tongue-zone and outer-tongue-zone; (b) The inner tongue zone moves with the position of the inner tongue; (c) A brief illustration of the reason of the two high current zones.

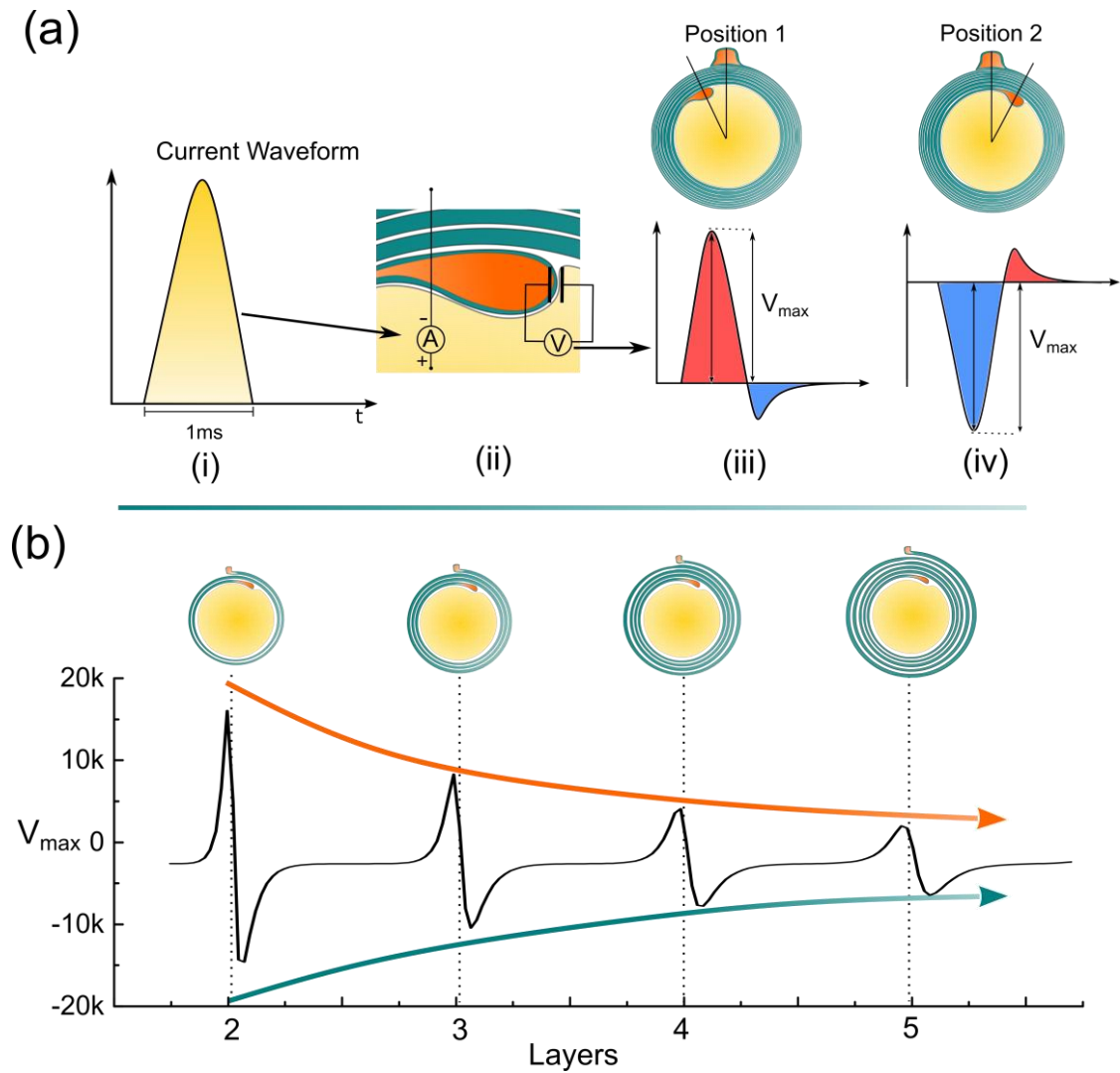

Figure 4 Simulation details and the measured voltage amplitude of the inner tongue. (a) i: A monophasic positive current pulse is used to model the action potential; ii: The detailed configuration of the applied current source and how the voltage on the inner tongue is measured; iii: The illustrative voltage waveform on the inner tongue when the inner tongue is at position 1; iv: The illustrative voltage waveform on the inner tongue when the inner tongue is at position 2; (b) The amplitude of the measured voltage amplitude (Vmax) by increasing the number of myelin layers.

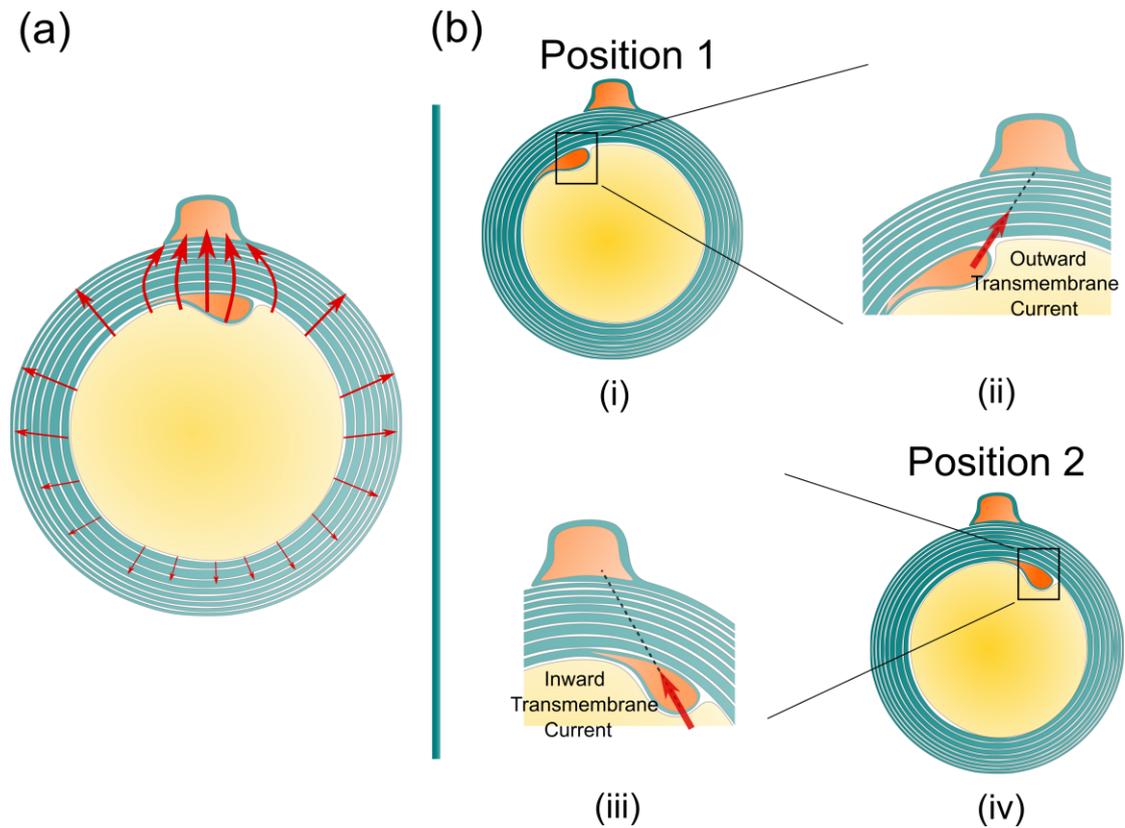

Figure 5 The illustrative current flow on the cross-section of the myelin sheath. (a) The current from the inner tends to flow towards the outer tongue; (b) A illustrative drawing explains the reason for the polarity reverse. i: when the inner tongue is at position 1, the current from the inner tongue to the outer tongue forms an outward transmembrane current on the inner tongue; ii: when the inner tongue is at position 2, the current from the inner tongue to the outer tongue forms an inward transmembrane current on the inner tongue.

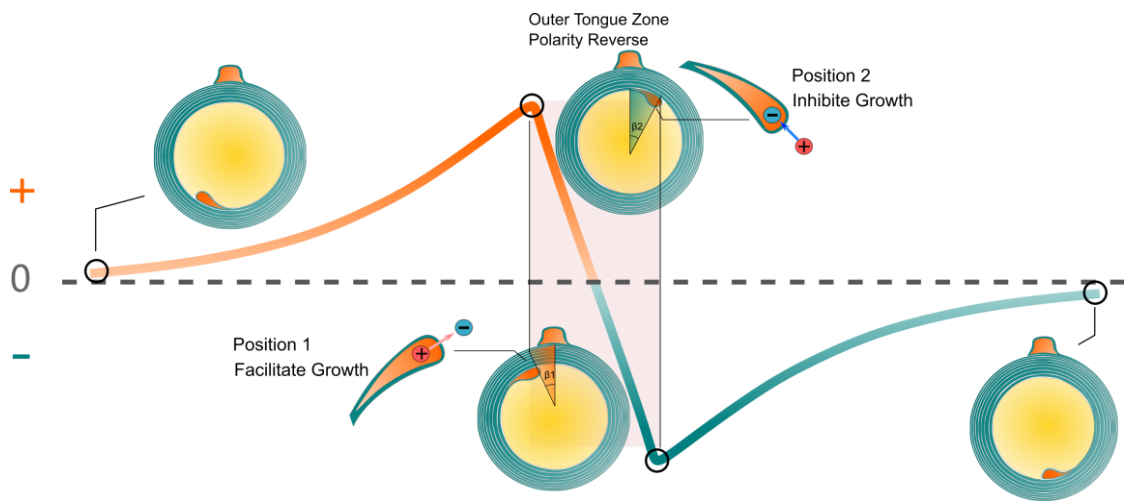

Figure 6. The relationship between the transmembrane voltage of the inner tongue and its radial position. The growth rate is correlated with the polarity (direction) and amplitude of the transmembrane E-field.

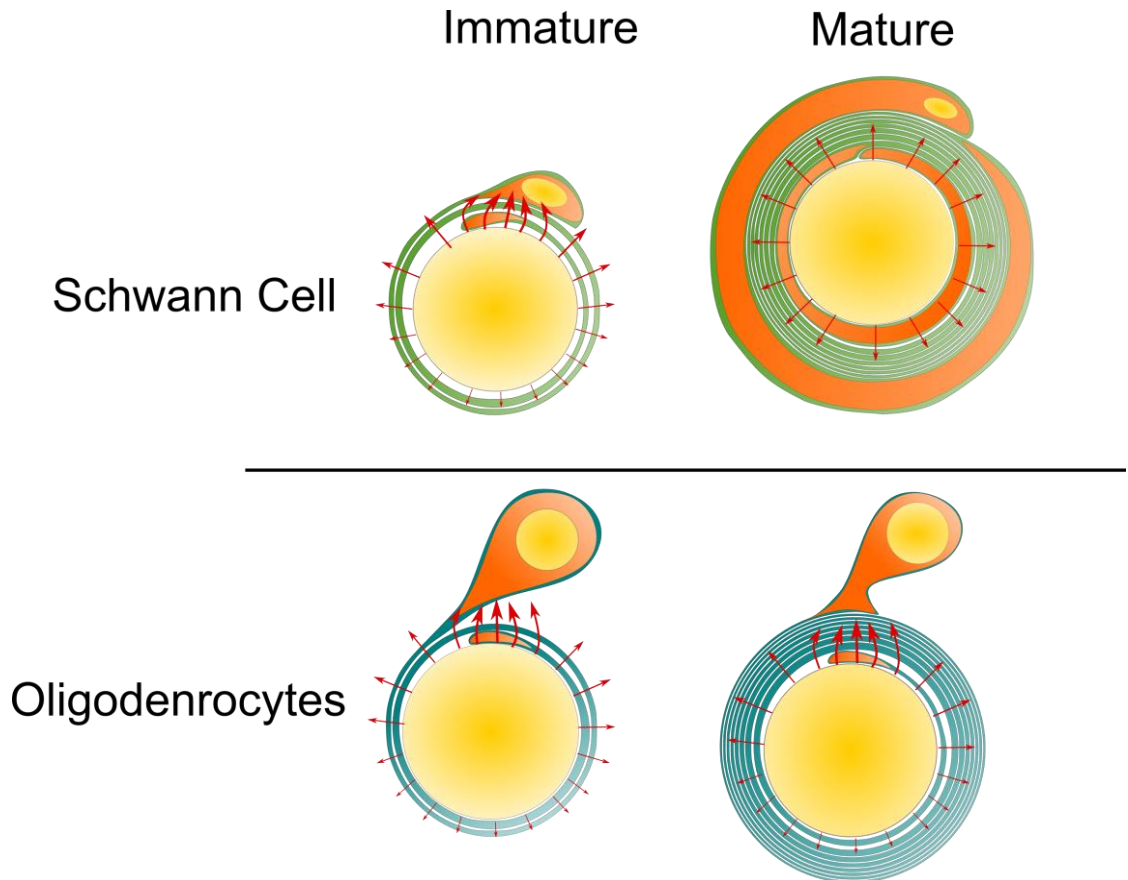

Figure 7. The same quadrant phenomenon for Schwann cell and Oligodendrocyte. For the Schwann cell, at its mature state, its inner tongue and outer tongue form two circles, eliminating the occurrence of the two high-current zones. So the same quadrant phenomenon tends to disappear with growth.

# Supplementary information of
# A physical perspective to understand myelin. I. A physical answer to Peter's quadrant mystery

**Authors**: Yonghong Liu[1], Yapeng Zhang[1], Wenji Yue[1], Ran Zhu[1], Tianruo Guo[3], Fenglin Liu[1], Yubin Huang[1], Tianzhun Wu*[1,2], Hao Wang*[1,2]

**Affiliations:**

[1]Institute of Biomedical & Health Engineering, Shenzhen Institutes of Advanced Technology (SIAT), Chinese Academy of Sciences (CAS), Shenzhen 518035, China

[2]Key Laboratory of Health Bioinformatics, Chinese Academy of Sciences

[3]Graduate School of Biomedical Engineering, University of New South Wales, Sydney, NSW2052, Australia

***Hao Wang** hao.wang@siat.ac.cn
***Tianzhun Wu** tz.wu@siat.ac.cn

## S1 The detailed circuit configuration of the modeling.

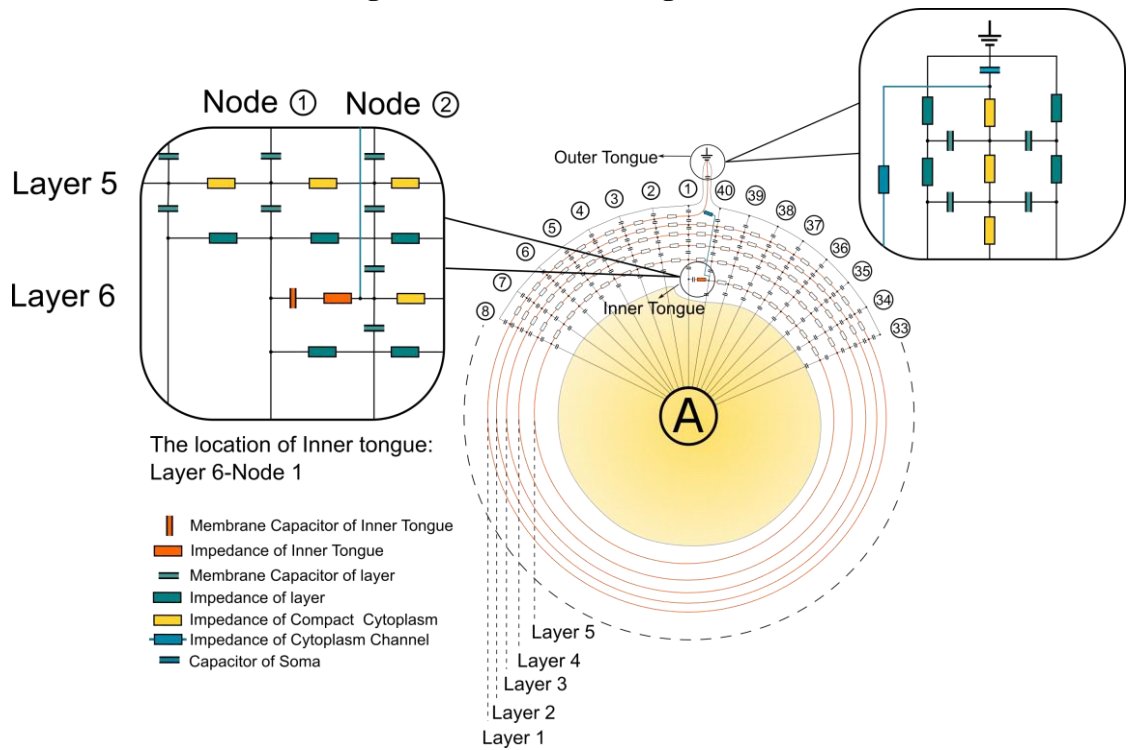

Figure S1 The detailed circuit configuration for modeling.

Figure S1 shows the circuit network used for the modeling. The actual parameters of each unit are based on the histological characteristics of myelin. The outer tongue is always located at Layer 1. There are 40 nodes for each layer. Then the growth progress of the myelin can be modeled by changing the position of the inner tongue. Thus,

For the modeling of dynamic progress of myelin growth, we just move the position of the inner tongue. As showed in Figure S1, the inner tongue is located at Layer 6-Node 1, meaning the inner tongue is overlap with the outer tongue at this status, and it is the first unit of Layer 6. At the same time, it was grown from Layer 5-Node 40, the previous state, and will grow to Layer 6-Node 2, in the next state. The voltage change progress of inner tongue's membrane capacitor is simulated by all different states of myelin growth progress, from layer 2 to layer 6.

## S2 The effect of current concentration at high-current-zone by cytoplasmic channel

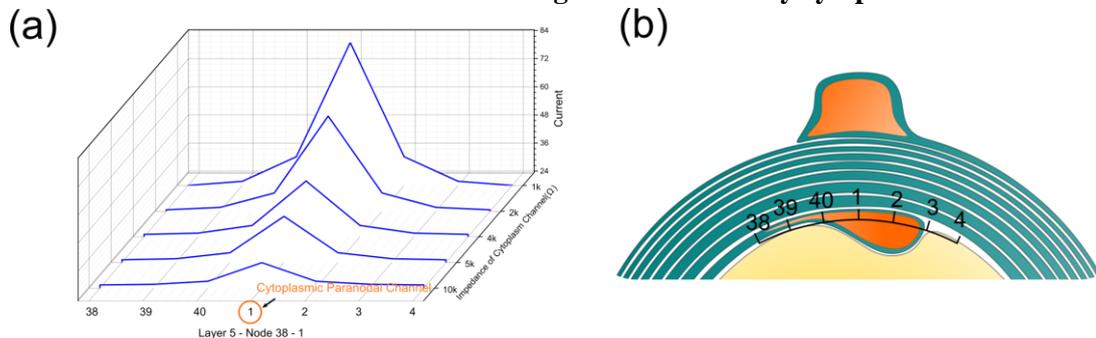

Figure S2. (a) illustrates the current concentration effect of cytoplasmic paranodal channel. The resistance represents the cytoplasmic channel is connected to Layer 5-Node 1. Then the current on the capacitors of peripheral region (Layer 5-Node 38,39,40,1,2,3,4) is measured by changing the resistance of the cytoplasmic channel. As seen, no matter how high is the impedance of the channel, as long as it exists, the capacitor closer to the channel always has higher current than peripheral region.

Table 1. Electrical Parameters of Elements in the myelin circuit network

| Elements | Parameter |
|---|---|
|  Membrane Capacitor of Inner Tongue | 50nF |
|  Impedance of Inner Tongue | 4Ω |
|  Membrane Capacitor of layer | 1nF |
|  Impedance of layer | 1kΩ |
|  Impedance of Compact Cytoplasm | 4kΩ |
|  Impedance of Cytoplasm Channel | 30MΩ |
|  Capacitor of Soma | 1mF |

*The parameters provided in this table correspond to the model to which this article belongs, and can be flexibly modified within the allowable range.